# General Physical Properties of Gamma-Ray-emitting Radio Galaxies

Yongyun Chen (陈永云)[1] , Qiusheng Gu (顾秋生)[2], Junhui Fan (樊军辉)[3], Xiaoling Yu (俞效龄)[1], Nan Ding (丁楠)[4],
Dingrong Xiong (熊定荣)[5], and Xiaotong Guo (郭晓通)[6]
[1] College of Physics and Electronic Engineering, Qujing Normal University, Qujing 655011, People's Republic of China; ynkmcyy@yeah.net
[2] School of Astronomy and Space Science, Nanjing University, Nanjing 210093, People's Republic of China; qsgu@nju.edu.cn
[3] Center for Astrophysics, Guangzhou University, Guangzhou 510006, People's Republic of China
[4] School of Physical Science and Technology, Kunming University, Kunming 650214, People's Republic of China
[5] Yunnan Observatories, Chinese Academy of Sciences, Kunming 650011, People's Republic of China
[6] Anqing Normal University, Anqing 246133, People's Republic of China
Received 2023 February 10; revised 2023 March 16; accepted 2023 March 17; published 2023 April 11

## Abstract

We study the radio galaxies with known redshifts detected by the Fermi satellite after 10 yr of data (4FGL-DR2). We use a one-zone leptonic model to fit the quasi-simultaneous multiwavelength data of these radio galaxies and study the distributions of the derived physical parameters as a function of black hole mass and accretion disk luminosity. The main results are as follows. (1) We find that the jet kinetic power of most radio galaxies can be explained by the hybrid jet model based on ADAFs surrounding Kerr black holes. (2) After excluding the redshift, there is a significant correlation between the radiation jet power and the accretion disk luminosity, while the jet kinetic power is weakly correlated with the accretion disk luminosity. (3) We also find a significant correlation between inverse Compton luminosity and synchrotron luminosity. The slope of the correlation for radio galaxies is consistent with the synchrotron self-Compton (SSC) process. The result may suggest that the high-energy components of radio galaxies are dominated by the SSC process.

*Unified Astronomy Thesaurus concepts:* Active galactic nuclei (16); Gamma-rays (637); Gamma-ray sources (633); Radio galaxies (1343); Galaxy jets (601)

## 1. Introduction

The Fermi Large Area Telescope (LAT) has detected that many active galactic nuclei (AGNs) have high-energy gamma-ray radiation, such as blazars, indicating that these AGNs have strong relativistic beaming jets. Although the number is small, Fermi-LAT also detected significant gamma-ray emissions from AGNs, which are misaligned, mainly radio galaxies (Ajello et al. 2020). However, they are being extensively studied and investigated to identify new gamma-ray emitters to study the radiation processes that power relativistic jets, as well as general AGN unification (e.g., Kataoka et al. 2011; Grandi et al. 2012; Paliya 2021). Radio galaxies are thought to be the parent of the blazar: the blazar shares our line of sight, while the jets of radio galaxies point to a larger viewing angle. According to the radio morphology of radio galaxies, radio galaxies are divided into two subclasses: Fanaroff–Riley type Is (FR Is) show bright jets near the nucleus, while Fanaroff–Riley type IIs (FR IIs) show prominent hotspots far away from the nucleus (Fanaroff & Riley 1974). This classification is also reflected in the separation of radio power (below and above $L_{1.4\,\mathrm{GHz}} = 10^{25}$ W Hz$^{-1}$, respectively). Fermi-LAT detected more FR Is than FR IIs. This may be because the gamma-ray spectrum of the former is relatively flat and/or due to their closer proximity compared to FR IIs. These two types are considered to be the misaligned counterparts of flat-spectrum radio quasars and BL Lacertae objects (BL Lacs), respectively. Most radio galaxies with gamma-ray emissions are found in the low-redshift universe ($z < 0.5$; Abdo et al. 2010; Chiaro et al. 2020). For example, in the second data release of the fourth catalog of AGNs detected by Fermi-LAT (4LAC-DR2; Ajello et al. 2020), the farthest known radio galaxy is 3C 17 = 0.22. Foschini et al. (2022) have identified that TXS 1433+205 is a misaligned AGN. Recently, Paliya et al. (2023) identified the most distant gamma-ray-emitting FR II radio galaxy, TXS 1433+205, at $z = 0.748$.

At present, the reason for the morphological differences between FR I and FR II radio galaxies is not clear. According to the theoretical model, there are two possible reasons: the morphological differences originate from: (1) different physical conditions in their ambient media (see Gopal-Krishna & Wiita 2000); and (2) the inherent differences in their central engines, that is, different accretion modes and/or jet formation processes (e.g., Bicknell 1995; Reynolds et al. 1996; Meier 1999; Ghisellini & Celotti 2001; Marchesini et al. 2004; Hardcastle et al. 2007; Wu & Cao 2008; Chen et al. 2015). The formation of jets has always been a hot issue in astrophysics. There are three main jet formation mechanisms. The first is the Blandford–Znajek (BZ) mechanism (Blandford & Znajek 1977), where jets extract the rotational energy of black holes. The second is the Blandford–Payne (BP) mechanism (Blandford & Payne 1982), in which the jet extracts the rotational energy of the accretion disk. In both cases, it should be maintained by the accretion of matter on the black hole, which leads to the expectation of the relationship between accretion and jet power (Maraschi & Tavecchio 2003). The third is the hybrid jet model, which is a combination of BP and BZ (Meier 2001; Garofalo et al. 2010).

Since the average position of radio galaxies is closer than that of blazars, it is possible to obtain more detailed observation results to test the theoretical model of their emissions. The broadband spectral energy distributions (SEDs) of AGN jets are very prominent in the emissions of blazars and radio galaxies and are nonthermal globally, which is characterized by the existence of two components (usually called "peak"; Blandford et al. 2019), namely the synchrotron component and the inverse Compton (IC) component (Ghisellini et al. 1997; Massaro et al. 2004, 2006). In the lepton model, the low-energy component is usually attributed to synchrotron emission, which is produced by the motion of relativistic particles in the presence of a magnetic field (Rybicki & Lightman 1979). The high-energy component of broadband SEDs is interpreted as the IC emission generated by the electron population in the jet (Massaro et al. 2004, 2006; Longair 2011). However, there are differences in the origins of the soft photons scattered by IC. One is that they come from synchrotron emission, namely the synchrotron self-Compton (SSC) process (Rees 1967; Jones et al. 1974; Marscher & Gear 1985; Maraschi et al. 1992; Sikora et al. 1994; Bloom & Marscher 1996). The other is that they come from the exteriors of jets, namely the external Compton (EC) process. There are three possible sources of EC soft photons: accretion disk photons directly entering the jets (Dermer et al. 1992; Dermer & Schlickeiser 1993); broad-line region (BLR) photons entering the jets (Sikora et al. 1994; Dermer et al. 1997); and dust torus infrared radiation photons entering the jets (Błażejowski et al. 2000; Arbeiter et al. 2002). Ghisellini (1996) derived two relations between synchrotron luminosity and IC luminosity, which can determine whether the IC component is dominated by the EC process or the SSC process ($L_{\rm EC} \sim L_{\rm syn}^{1.5}$, $L_{\rm SSC} \sim L_{\rm syn}^{1.0}$). The broadband emission of M87 is usually explained by the one-zone SSC scenario (Abdo et al. 2009; de Jong et al. 2015).

In this work, we try to understand the physical properties of these radio galaxies with gamma-ray emission. We use a one-zone SSC jet model to fit the quasi-simultaneous multi-wavelength data of radio galaxies with gamma-ray emission and some physical parameters are obtained, such as the jet power of the radiation, electrons, magnetic field, synchrotron peak frequency luminosity, IC luminosity, and so on. The samples are presented in Section 2. In Section 3, we present the jet model. In Section 4, we describe the results and discussions. Section 5 presents the conclusions.

## 2. The Sample

### 2.1. The Radio Galaxy Sample

We get radio galaxy samples from the data released in the 4FGL-DR2 catalog. Fermi-LAT has detected 41 radio galaxies (Ajello et al. 2020). We carefully examine the sample of 41 radio galaxies and compare it with the source classifications of Foschini et al. (2022) and Angioni (2020). According to the source classifications of Foschini et al. (2022) and Angioni (2020), these radio galaxies are mainly divided into two types: FR Is and FR IIs. We only select radio galaxies with reliable redshifts, 1.4 GHz radio flux, and the absolute magnitude ($M_{\rm H}$) of the H band. Since there are almost no FR0 radio galaxies in our sample, we do not consider FR0 radio galaxies. We derive 38 radio galaxies (nine FR II radio galaxies and 29 FR I radio galaxies). We get the radio flux of 1.4 GHz and the absolute magnitude of the H band of these radio galaxies from the NASA/IPAC Extragalactic Database. The 1.4 GHz radio luminosity is estimated by using $L_{\rm radio} = 4\pi d_{\rm L}^2(1+z)^{a-1}\nu S_\nu$, where $S_\nu$ is the flux density, $z$ is the redshift, and $d_L$ is the the luminosity distance. The radio spectral index $a = 0$ is used (Abdo et al. 2010). The mass of the black hole is estimated using the following formula (Marconi & Hunt 2003; Sbarrato et al. 2014):

$$\log(M_{\rm BH}/M_\odot) = -2.77 - 0.464 \times M_{\rm H}. \quad (1)$$

### 2.2. The Jet Power

We use the one-zone leptonic model (SSC; Tramacere et al. 2009, 2011; Tramacere 2020) to fit the quasi-simultaneous multiwavelength data of the radio galaxies. The quasi-simultaneous multiwavelength data of the radio galaxies come from the Space Science Data Center SED Builder.[7] The broadband SEDs have been modeled using the JetSet[8] numerical leptonic code (Tramacere 2020). The parameters with the minimum $\chi^2$/degrees of freedom were defined as the best-fit values. We get the jet power of the Poynting flux, radiation, electrons, and protons (assuming one proton per emitting electron). The jet power in different forms is calculated from

$$P_{\rm i} = \pi R^2 \Gamma^2 \beta c U'_{\rm i}, \quad (2)$$

where $U'_{\rm i}$ is the energy density of the protons ($i = p$), the relativistic electrons ($i = e$), the magnetic field ($i = B$), and the produced radiation ($i = {\rm rad}$). The jet power of radiation, $P_{\rm r} = \pi R^2 \Gamma^2 \beta c U'_{\rm rad}$, can be estimated by using the radiation energy density ($U'_{\rm rad} = L'/(4\pi R^2 c)$):

$$P_{\rm r} = L'\frac{\Gamma^2}{4} = L\frac{\Gamma^2}{4\delta^4} \sim L\frac{1}{4\delta^2}, \quad (3)$$

where $L$ is the total observed nonthermal luminosity ($L'$ is in the comoving frame). $\delta$ is the Doppler factor. The relevant data are shown in Table 1. An example is shown in Figure 1.

## 3. The Jet Model

The most popular jet formation mechanisms at present include the BZ process (Blandford & Znajek 1977) and the BP process (Blandford & Payne 1982). Recently, a hybrid jet model has been proposed by many authors (Meier 2001; Garofalo et al. 2010), namely a mixture of BZ and BP. Most previous works on the jet power extracted from advection-dominated accretion flows (ADAFs) have been based on self-similar solutions of ADAFs (e.g., Narayan & Yi 1995; Meier 2001; Nemmen et al. 2007). There are several pieces of evidence suggesting that the accretion flows on AGNs with low accretion rates (or low jet power, such as BL Lacs) are best described as ADAFs. Moreover, the magnitude and the structure of the magnetic fields associated with ADAFs are much more conducive to the extraction of spin energy from the black hole than are those associated with a standard thin disk (e.g., Livio et al. 1999; Nemmen et al. 2007). At the same time, we find that our sample has low accretion rates (see Figure 4), which implies that these sources may have ADAFs. In this work, therefore, we calculate the jet power based on the self-similar solutions of ADAFs surrounding Kerr black holes.

(1) The BZ jet model.

According to the BZ model, the jet power is given by (e.g., MacDonald & Thorne 1982; Thorne et al. 1986; Ghosh &

---

[7] http://tools.ssdc.asi.it/SED/
[8] https://jetset.readthedocs.io/en/latest/

Table 1
The Sample of Gamma-Ray-emitting Radio Galaxies

| Name (1) | R.A. (2) | Decl. (3) | Redshift (4) | $\log(M/M_\odot)$ (5) | $\log L_\gamma$ (6) | $B$ (7) | $\delta$ (8) | $\log P_{rad}$ (9) | $\log P_e$ (10) | $\log P_B$ (11) | $\log P_p$ (12) | $\log L_{syn}$ (13) | $\log L_{IC}$ (14) | Type (15) | Reference (16) |
|---|---|---|---|---|---|---|---|---|---|---|---|---|---|---|---|
| IC1531 | 2.4 | −32.38 | 0.026 | 8.811 | 42.56 | 0.361 | 5.01 | 40.77 | 42.50 | 43.29 | 42.53 | 43.11 | 41.28 | FRI | Fo22 |
| 3C 17 | 9.6899 | −2.0722 | 0.221 | 8.742 | 44.85 | 0.0297 | 24.53 | 39.46 | 42.95 | 41.95 | 43.10 | 44.21 | 44.98 | FRII | Fo22 |
| NGC 315 | 14.4498 | 30.397 | 0.016 | 9.187 | 42.35 | 0.04 | 32.21 | 40.36 | 43.92 | 40.72 | 43.45 | 43.16 | 42.45 | FRI | Fo22 |
| TXS 0149+710 | 28.3628 | 71.2411 | 0.023 | 8.746 | 42.66 | 2.06 | 49.47 | 43.40 | 42.57 | 44.80 | 40.42 | 43.49 | 42.78 | FRI | An20 |
| PKS 0235+017 | 39.4414 | 2.1019 | 0.022 | 8.719 | 42.23 | 3.175 | 31.62 | 40.68 | 42.17 | 41.66 | 41.28 | 43.41 | 42.74 | FRI | An20 |
| NGC 1218 | 47.1105 | 4.1177 | 0.028 | 9.331 | 43.14 | 0.059 | 21.88 | 40.79 | 43.98 | 41.45 | 44.17 | 43.30 | 42.95 | FRI | Fo22 |
| B3 0309+411B | 48.2438 | 41.3289 | 0.134 | 9.758 | 44.38 | 0.00138 | 34.67 | 40.67 | 45.59 | 39.58 | 44.68 | 44.56 | 44.69 | FRII | Fo22 |
| IC 310 | 49.2147 | 41.3464 | 0.019 | 8.844 | 42.36 | 0.000647 | 49.99 | 39.89 | 45.41 | 40.59 | 45.73 | 43.89 | 42.32 | FRI | An20 |
| NGC 1275 | 49.9575 | 41.5121 | 0.0176 | 9.155 | 44.28 | 0.081 | 22.63 | 41.85 | 44.01 | 42.76 | 43.30 | 43.66 | 44.14 | FRI | Fo22 |
| Fornax A | 50.67 | −37.21 | 0.006 | 9.382 | 41.68 | 0.0115 | 12.48 | 40.80 | 43.59 | 40.81 | 42.95 | 42.30 | 41.59 | FRI | An20 |
| 3C 111 | 64.5633 | 38.1219 | 0.049 | 8.862 | 43.98 | 0.00955 | 36.87 | 41.96 | 45.84 | 40.54 | 45.63 | 43.86 | 43.47 | FRII | An20 |
| 3C 120 | 68.2618 | 5.3696 | 0.033 | 8.575 | 43.53 | 0.064 | 23.49 | 42.44 | 44.69 | 42.48 | 43.85 | 43.85 | 43.10 | FRI | Fo22 |
| Pictor A | 79.9058 | −45.7439 | 0.035 | 7.897 | 43.13 | 0.022 | 18.44 | 41.71 | 44.77 | 41.63 | 43.86 | 43.46 | 43.64 | FRII | Fo22 |
| PKS 0625-35 | 96.7747 | −35.4867 | 0.055 | 9.197 | 44.08 | 0.0082 | 47.59 | 40.89 | 44.67 | 41.16 | 44.01 | 43.91 | 43.46 | FRI | Fo22 |
| NGC 2329 | 107.2472 | 48.6558 | 0.019 | 8.895 | 42.01 | 0.168 | 49.27 | 41.01 | 44.53 | 44.26 | 45.34 | 43.11 | 42.61 | FRI | Fo22 |
| NGC 2484 | 119.6893 | 37.7776 | 0.041 | 9.266 | 42.76 | 0.0216 | 23.42 | 41.27 | 43.83 | 42.61 | 42.78 | 43.86 | 42.32 | FRI | Fo22 |
| B2 1113+29 | 169.1625 | 29.2586 | 0.049 | 8.784 | 42.59 | 0.165 | 19.96 | 41.14 | 43.14 | 41.61 | 41.70 | 43.99 | 43.47 | FRI | Fo22 |
| 3C 264 | 176.2394 | 19.6283 | 0.0216 | 8.849 | 42.53 | 0.077 | 39.96 | 39.78 | 43.65 | 40.47 | 43.78 | 43.33 | 41.95 | FRI | Fo22 |
| NGC 3894 | 177.2539 | 59.4158 | 0.011 | 8.612 | 41.80 | 0.099 | 25.37 | 40.57 | 43.31 | 41.49 | 42.79 | 43.57 | 42.12 | FRI | An20 |
| M 87 | 187.7123 | 12.3883 | 0.004 | 7.048 | 41.76 | 0.052 | 13.31 | 40.35 | 43.11 | 41.47 | 42.66 | 43.05 | 41.91 | FRI | Fo22 |
| PKS 1234-723 | 189.237 | −72.5336 | 0.024 | 8.890 | 42.58 | 0.089 | 23.11 | 40.95 | 43.51 | 40.59 | 42.05 | 43.42 | 42.81 | FRI | |
| TXS 1303+114 | 196.5992 | 11.229 | 0.086 | 9.030 | 43.53 | 0.068 | 42.81 | 41.54 | 44.28 | 39.89 | 42.86 | 45.03 | 44.00 | FRI | Fo22 |
| PKS 1304-215 | 196.6988 | −21.8094 | 0.126 | 8.983 | 44.44 | 0.034 | 20.93 | 40.71 | 43.42 | 39.75 | 41.73 | 42.91 | 44.14 | FRII | An20 |
| Cen A | 201.37 | −43.02 | 0.00183 | 8.185 | 41.64 | 0.00024 | 43.84 | 39.56 | 44.68 | 41.03 | 42.81 | 41.86 | 41.01 | FRI | Fo22 |
| Cen B | 206.5988 | 72.4612 | 0.01292 | 8.969 | 41.53 | 0.096 | 7.15 | 39.88 | 42.38 | 40.12 | 40.97 | 41.91 | 41.73 | FRI | An20 |
| 3C 303 | 220.7789 | 52.0294 | 0.141 | 8.992 | 43.97 | 0.032 | 22.87 | 41.73 | 45.92 | 41.90 | 46.24 | 43.90 | 43.57 | FRII | Fo22 |
| B2 1447+27 | 222.3956 | 27.7686 | 0.031 | 8.862 | 42.30 | 0.038 | 22.19 | 40.66 | 44.06 | 40.33 | 43.91 | 43.95 | 43.05 | FRI | Fo22 |
| PKS 1514+00 | 229.1375 | 0.2657 | 0.053 | 8.849 | 43.50 | 0.0182 | 15.69 | 41.60 | 44.96 | 41.05 | 45.32 | 43.87 | 43.54 | FRII | Fo22 |
| TXS 1516+064 | 229.6503 | 6.2419 | 0.102 | 9.178 | 43.67 | 0.042 | 37.12 | 41.02 | 44.00 | 40.84 | 42.67 | 44.30 | 44.17 | FRI | Fo22 |
| PKS B1518+045 | 230.2949 | 4.3644 | 0.052 | 8.895 | 43.06 | 0.266 | 11.62 | 42.36 | 43.53 | 41.14 | 42.07 | 43.77 | 43.52 | FRI | Fo22 |
| NGC 6251 | 247.6686 | 82.5742 | 0.0247 | 9.132 | 43.24 | 0.039 | 24.6 | 40.83 | 44.24 | 40.84 | 43.58 | 43.36 | 43.12 | FRI | Fo22 |
| NGC 6328 | 261.0666 | −65.0174 | 0.0144 | 8.654 | 42.09 | 0.378 | 38.08 | 40.49 | 42.73 | 41.78 | 41.73 | 43.45 | 42.36 | FRII | Fo22 |
| PKS 1839-48 | 280.8674 | −48.5913 | 0.111 | 9.062 | 43.74 | 0.202 | 11.65 | 43.23 | 44.26 | 40.23 | 43.08 | 44.79 | 44.25 | FRI | Fo22 |
| PKS 2153-69 | 329.025 | −69.7067 | 0.0283 | 8.784 | 42.82 | 0.193 | 9.25 | 41.94 | 43.84 | 42.15 | 43.28 | 43.56 | 43.08 | FRI | Fo22 |
| PKS 2225-308 | 336.981 | −30.5174 | 0.058 | 9.405 | 43.09 | 0.035 | 49.91 | 40.36 | 43.78 | 40.47 | 40.21 | 43.61 | 43.52 | FRI | Fo22 |
| PKS 2300-18 | 345.7152 | −18.6985 | 0.129 | 8.821 | 44.07 | 0.079 | 15.56 | 42.15 | 44.43 | 41.50 | 43.73 | 44.19 | 44.37 | FRI | Fo22 |
| PKS 2324-02 | 351.7355 | −2.023 | 0.189 | 9.415 | 44.65 | 0.033 | 27.92 | 40.70 | 45.45 | 41.04 | 46.33 | 44.30 | 44.81 | FRII | Fo22 |
| PKS 2327-215 | 352.4375 | −21.3035 | 0.031 | 8.408 | 43.02 | 0.055 | 26.13 | 40.87 | 43.63 | 40.48 | 42.08 | 42.83 | 43.05 | FRI | Fo22 |

**Note.** Column (1): the name of the source. Column (2): the right ascension in decimal degrees. Column (3): the decl. in decimal degrees. Column (4): the redshift. Column (5): the black hole mass. Column (6): the gamma-ray luminosity. Column (7): the magnetic field, in units of gauss. Column (8): the Doppler factor. Column (9): the radiation jet power, in units of ergs per second. Column (10): the electron jet power, in units of ergs per second. Column (11): the magnetic field jet power, in units of ergs per second. Column (12): the proton jet power, in units of ergs per second. Column (13): the synchrotron peak frequency luminosity, in units of ergs per second. Column (14): the IC luminosity, in units of ergs per second. Column (15): the type. Column (16): the reference for the type—Fo22: Foschini et al. (2022); An20: Angioni (2020). PKS 1234–723 is classified by the method of Angioni (2020), namely by using the 1.4 GHz radio luminosity ($L_{1.4GHz} = 10^{25}$W/Hz).

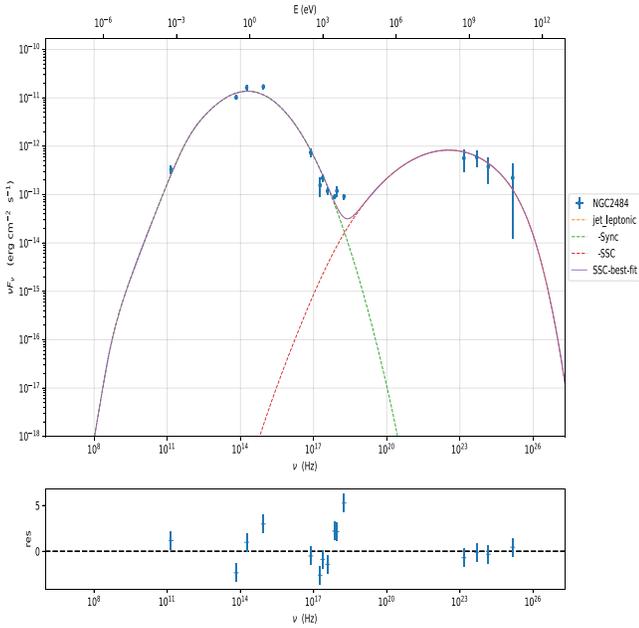

**Figure 1.** An example of a broadband SED of NGC 2484, modeled by using a one-zone model.

Abramowicz 1997; Nemmen et al. 2007):

$$P_{\text{jet}}^{\text{BZ}} = \frac{1}{32}\omega_F^2 B_\perp^2 R_H^2 j^2 c, \quad (4)$$

where $R_H = [1 + (1 - j^2)^{1/2}]GM/c^2$ is the horizon radius, $B_\perp$ is assumed to approximate to the poloidal component $B_p$, $B_\perp \approx B_p(R_{\text{ms}}) \approx g(R_{\text{ms}})B(R_{\text{ms}})$ (Livio et al. 1999), and $g = \Omega/\Omega'$. The $\omega_F \equiv \Omega_F(\Omega_H - \Omega_F)/\Omega_H^2$ depends on the angular velocity of the field lines $\Omega_F$ relative to that of the hole $\Omega_H$. To estimate the maximal power extracted from a spinning black hole, we assume that $\omega_F = 1/2$ (e.g., MacDonald & Thorne 1982; Thorne et al. 1986; Nemmen et al. 2007).

(2) The hybrid jet model.

Contrary to the premise of extremely low plasma density, based on the BZ model, the accretion of the matter is considered to be a key element of the real AGN system. In fact, numerical simulations show that the coupling between the accretion flow and the magnetic field is an important factor in jet formation. Therefore, we consider the hybrid jet model of Meier (2001). Meier (2001) has suggested that the jet power of the hybrid model strongly depends on the thickness of the accretion disk, the spin of the black hole, the intensity of the magnetic field, and rapid rotation. In this model, the magnetic fields extract energy from both the accretion flow and the spinning hole (Meier 2001). The jet power of the hybrid model is given by Meier (2001) and Nemmen et al. (2007):

$$P_{\text{jet}}^{\text{Hybrid}} = (B_\phi H R \Omega)^2 / 32c, \quad (5)$$

where $B_\phi = gB$ and $\Omega = \Omega' + \omega$. All quantities are evaluated at $R = R_{\text{ms}}$, which is also assumed to be the approximate characteristic size of the jet formation region.

Narayan & Yi (1995) described the self-similar ADAF structure, with the black hole mass in solar units ($m = M_{\text{BH}}/M_\odot$), accretion rates in Eddington units ($\dot{m} = \dot{M}/\dot{M}_{\text{Edd}}$, $\dot{M}_{\text{Edd}}$ is the Eddington accretion rate, $\dot{M}_{\text{Edd}} \equiv 22 M_{\text{BH}}/(10^9 M_\odot) M_\odot \text{ yr}^{-1}$), and radii in Schwarzschild units $[r = R/(2GM_{\text{BH}}/c^2)]$:

$$\Omega' = 7.19 \times 10^4 c_2 m^{-1} r^{-3/2} s^{-1}, \quad (6)$$

$$B = 6.55 \times 10^8 \alpha^{-1/2}(1 - \beta)^{1/2} c_1^{-1/2} c_3^{1/4} m^{-1/2} \dot{m}^{1/2} r^{-5/4} G, \quad (7)$$

$$H/R \approx (2.5 c_3)^{1/2}. \quad (8)$$

The constants $c_1$, $c_2$, and $c_3$ are defined as:

$$c_1 = \frac{5 + 2\epsilon'}{3\alpha^2} g'(\alpha, \epsilon'),$$
$$c_2 = \left[\frac{2\epsilon'(5 + 2\epsilon')}{9\alpha^2} g'(\alpha, \epsilon')\right]^{1/2},$$
$$c_3 = c_2^2/\epsilon',$$
$$\epsilon' \equiv \frac{1}{f}\left(\frac{5/3 - \gamma}{\gamma - 1}\right),$$
$$g'(\alpha, \epsilon') \equiv \left[1 + \frac{18\alpha^2}{(5 + 2\epsilon')^2}\right]^{1/2}. \quad (9)$$

The relations among $\alpha$, $\beta$, and $\gamma$ are given as follows:

$$\gamma = (5\beta + 8)/3(2 + \beta),$$
$$\alpha \approx 0.55/(1 + \beta), \quad (10)$$

where $\alpha$ is the viscosity parameter and $\beta$ is the ratio of gas to magnetic pressure (Hawley et al. 1995; Esin et al. 1997). The angular velocity of the local metric is given by Bardeen et al. (1972),

$$\omega \equiv -\frac{g_{\phi t}}{g_{\phi\phi}} = \frac{2jM_{\text{BH}}}{j^2(R + 2M_{\text{BH}}) + R^3}, \quad (11)$$

using geometrized units ($G = c = 1$). The marginally stable orbit of the accretion disk ($R_{\text{ms}}$) is given by (Bardeen et al. 1972):

$$R_{\text{ms}} = GM_{\text{BH}}/c^2\{3 + Z_2 - [(3 - Z_1)(3 + Z_1 + 2Z_2)]^{1/2}\},$$
$$Z_1 \equiv 1 + (1 - j^2)^{1/3}[(1 + j)^{1/3} + (1 - j)^{1/3}],$$
$$Z_2 \equiv (3j^2 + Z_1^2)^{1/2}.$$
$$(12)$$

## 4. Results and Discussion

Figure 2 shows the distribution of the physical parameters obtained from the model fitting of the SEDs of the sample of radio galaxies. The red shaded areas are FR II radio galaxies, and the green shaded areas are FR I radio galaxies. We study the difference in their physical parameter distribution by using the parameter T-test, the nonparametric Kolmogorov–Smirnov (K-S) test, and the Kruskal–Wallis H-test. The parameter T-test is mainly used to test whether there is a difference in the mean values of the physical parameters of two independent samples. The nonparametric K-S test and Kruskal–Wallis H-test are mainly used to test whether there is a difference in the distribution of the physical parameters of two independent samples. We assume that there are differences in the three tests at the same time, so that there is a significant difference in the physical parameters between the two samples.

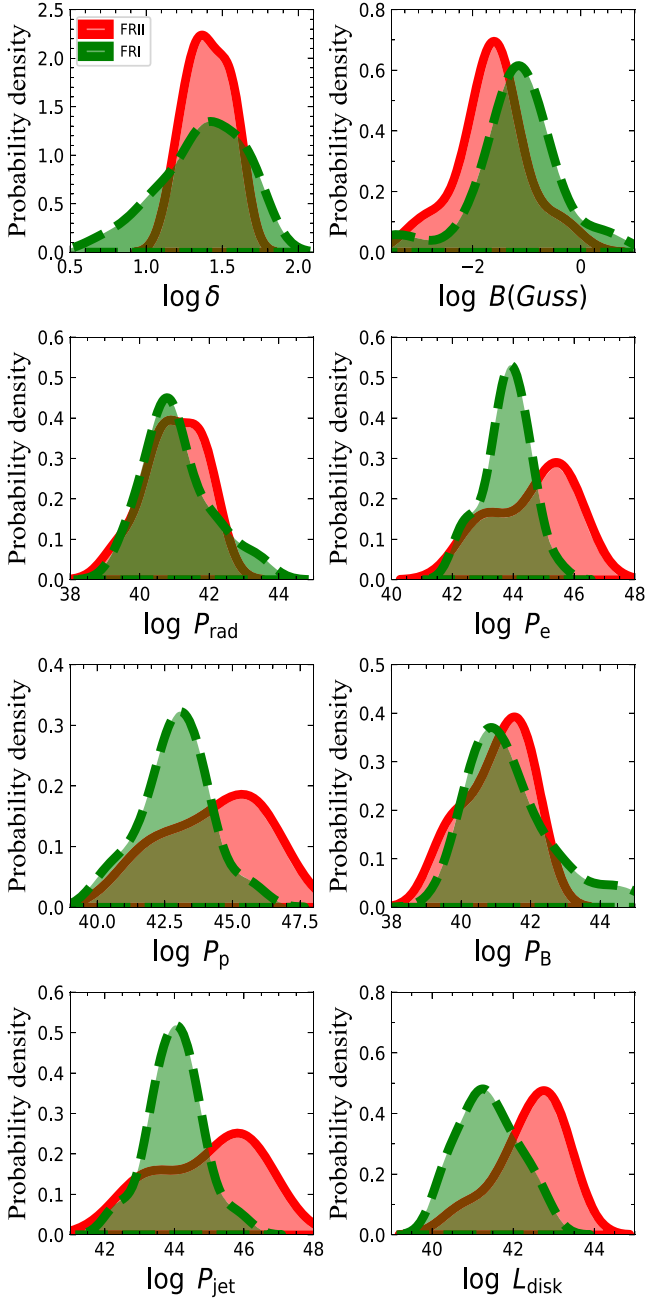

**Figure 2.** The distributions of physical parameters. The jet kinetic power is $P_{jet} = P_B + P_e + P_p$. The red line is for FR II radio galaxies and the green line is for FR I radio galaxies.

### 4.1. The Doppler Factor

The distribution range of the Doppler factor in FR I radio galaxies is larger than that in FR II radio galaxies (Figure 2). The average Doppler factors of FR II radio galaxies and FR I radio galaxies are $\langle \log \delta_{FRII} \rangle = 1.41 \pm 0.13$ and $\langle \log \delta_{FRI} \rangle = 1.37 \pm 0.26$, respectively. We find that the two averages are consistent with each other within the measurement errors. There is no significant difference in the average Doppler factor between FR I and FR II radio galaxies using the parameter T-test ($P = 0.67$; significant probability $P < 0.05$). Through a nonparametric K-S test ($P = 0.57$; significant probability $P < 0.05$) and a Kruskal–Wallis H-test ($P = 0.99$; significant probability $P < 0.05$), we also find that the distributions of the Doppler factor between FR II radio galaxies and FR I radio galaxies are not significantly different.

### 4.2. The Magnetic Field

The average magnetic fields of FR II radio galaxies and FR I radio galaxies are $\langle \log B_{FRII} \rangle = -1.63 \pm 0.60$ and $\langle \log B_{FRI} \rangle = -1.22 \pm 0.80$, respectively. The average magnetic fields of FR II radio galaxies and FR I radio galaxies are consistent with each other within the measurement errors. According to the parameter T-test ($P = 0.17$), the nonparametric K-S test ($P = 0.0006$), and the Kruskal–Wallis H-test ($P = 0.02$), we find that there is no significant difference in the distribution of the magnetic field between FR II radio galaxies and FR I radio galaxies.

### 4.3. The Jet Power

The average jet powers of radiation of FR II radio galaxies and FR I radio galaxies are $\langle \log P_{rad, FRII} \rangle = 41.00 \pm 0.76$ and $\langle \log P_{rad, FRI} \rangle = 41.08 \pm 0.94$, respectively. The average jet powers of radiation of FR II radio galaxies and FR I radio galaxies are consistent with each other within the measurement errors. The parameter T-test shows that there is no significant difference between these two averages ($P = 0.83$). Through a nonparametric K-S test ($P = 0.84$) and a Kruskal–Wallis H-test ($P = 0.88$), we find that the distributions of the jet power of radiation between FR II radio galaxies and FR I radio galaxies are not significantly different.

The average values of the jet power of electrons of FR II radio galaxies and FR I radio galaxies are $\langle \log P_{e, FRII} \rangle = 44.63 \pm 1.19$ and $\langle \log P_{e, FRI} \rangle = 43.78 \pm 0.74$, respectively. According to the parameter T-test ($P = 0.02$), nonparametric K-S test ($P = 0.004$), and Kruskal–Wallis H-test ($P = 0.052$), we find that the distributions of the jet power of electrons between FR II radio galaxies and FR I radio galaxies are not significantly different.

The average values of the jet power of protons of FR II radio galaxies and FR I radio galaxies are $\langle \log P_{p, FRII} \rangle = 44.29 \pm 1.69$ and $\langle \log P_{p, FRI} \rangle = 42.90 \pm 1.24$, respectively. The average jet powers of protons of FR II radio galaxies and FR I radio galaxies are consistent with each other within the measurement errors. Through a parameter T-test ($P = 0.01$), nonparametric K-S test ($P = 0.05$), and Kruskal–Wallis H-test ($P = 0.04$), we find that the distributions of the jet power of protons between FR II radio galaxies and FR I radio galaxies are not significantly different.

The average values of the jet power of the magnetic field of FR II radio galaxies and FR I radio galaxies are $\langle \log P_{B, FRII} \rangle = 41.02 \pm 0.85$ and $\langle \log P_{B, FRI} \rangle = 41.42 \pm 1.18$, respectively. The average jet powers of the magnetic fields of FR II radio galaxies and FR I radio galaxies are not significantly different within the measurement errors. Through a parameter T-test ($P = 0.37$), nonparametric K-S test ($P = 0.72$), and Kruskal–Wallis H-test ($P = 0.77$), we find that the distributions of the jet power of magnetic field between FR II radio galaxies and FR I radio galaxies are not significantly different.

The average values of the jet kinetic power of FR II radio galaxies and FR I radio galaxies are $\langle \log P_{jet, FRII} \rangle = 44.93 \pm 1.32$ and $\langle \log P_{jet, FRI} \rangle = 44.02 \pm 0.76$, respectively. The average jet kinetic powers of FR II radio galaxies and FR I radio galaxies are consistent with each other within the measurement errors. Through a parameter T-test ($P = 0.02$), nonparametric K-S test ($P = 0.008$), and Kruskal–Wallis H-test ($P = 0.08$), we find that the distributions of jet kinetic power between FR II radio galaxies and FR I radio galaxies are not significantly different.

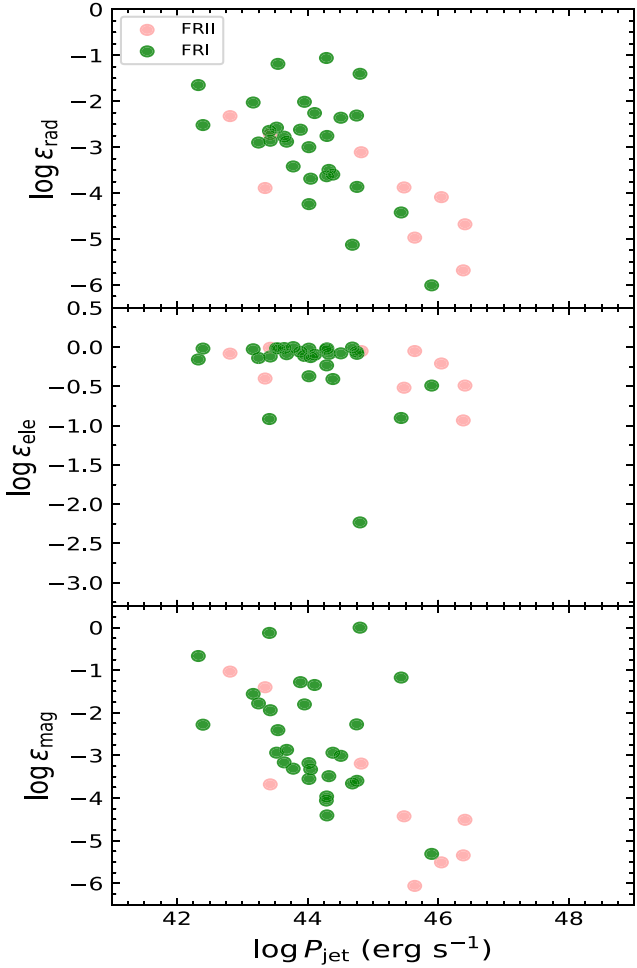

**Figure 3.** The fraction of the total jet power transformed into radiation (top), relativistic electrons (middle), and Poynting flux (bottom). The red dots show FR II radio galaxies and the green dots show FR I radio galaxies.

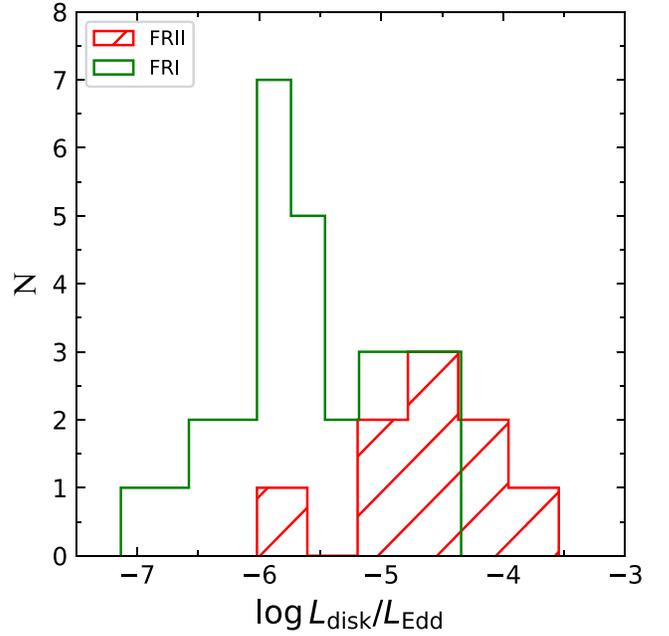

**Figure 4.** The distributions of the accretion rates of radio galaxies. The red line shows FR II radio galaxies and the green line shows FR I radio galaxies.

It is found that the gamma-ray luminosity is a good indicator of the BLR luminosity and that the following empirical relation holds (Ghisellini et al. 2010; Sbarrato et al. 2012; Paliya et al. 2017): $L_{BLR} \sim 4 L_\gamma^{0.93}$. Under the assumption that the BLR reprocesses 10% of the accretion disk luminosity, we derive the luminosity of the accretion disk. The average values of disk luminosity of FR II radio galaxies and FR I radio galaxies are $\langle \log L_{disk, FRII} \rangle = 42.42 \pm 0.76$ and $\langle \log L_{disk, FRI} \rangle = 41.39 \pm 0.71$, respectively. Through a parameter T-test ($P = 0.0008$), nonparametric K-S test ($P = 0.01$), and Kruskal–Wallis H-test ($P = 0.003$), we find that the distributions of accretion disk luminosity between FR II radio galaxies and FR I radio galaxies are significantly different.

In Figure 3, we show the fraction of jet kinetic power ($P_{jet} = P_e + P_p + P_B$) converted to radiation ($\epsilon_{rad}$), carried by relativistic electrons ($\epsilon_{ele}$) and the magnetic field ($\epsilon_{mag}$). The average values of $\epsilon_{rad}$ of FR II radio galaxies and FR I radio galaxies are $\langle \log \epsilon_{rad, FRII} \rangle = -3.93 \pm 1.03$ and $\langle \log \epsilon_{rad, FRI} \rangle = -2.94 \pm 1.09$, respectively. The average values of $\epsilon_{ele}$ of FR II radio galaxies and FR I radio galaxies are $\langle \log \epsilon_{ele, FRII} \rangle =$ $-0.30 \pm 0.29$ and $\langle \log \epsilon_{ele, FRI} \rangle = -0.24 \pm 0.44$, respectively. The average values of $\epsilon_{mag}$ of FR II radio galaxies and FR I radio galaxies are $\langle \log \epsilon_{mag, FRII} \rangle = -3.91 \pm 1.67$ and $\langle \log \epsilon_{mag, FRI} \rangle = -2.60 \pm 1.25$, respectively. From the above results, we find that FRI and FRII radio galaxies have $P_{rad} < P_e$. This may be a direct result of the slow cooling of the electrons. Its time is much longer than the time of light passing through the emission region, so the energy released in the form of radiation is less than the energy remaining in the electrons. The radio galaxies detected by gamma-rays have low accretion rates (see Figure 4), indicating that they are radiation-ineffective and that the relativistic electrons of the jets of FR I radio galaxies and FR II radio galaxies are slow-cooling (e.g., Sbarrato et al. 2014). The FR I radio galaxies and FR II radio galaxies have $P_{rad} \sim P_{mag}$, which may imply that the magnetic field alone is responsible for the observed radiation. At the same time, we also find that most radio galaxies have $\log \epsilon_{mag} < 0$, which implies that the jet kinetic powers of these radio galaxies are not dominated by the Poynting flux (Zdziarski et al. 2015). From the top panel of Figure 3, we can also see that all radio galaxies have $\log \epsilon_{rad} < 0$, which implies that the jet power of the kinetic energy is greater than that of radiation. Ghisellini et al. (2014) have also found that the jet kinetic power is larger than the radiation jet power for Fermi blazars. Our results are consistent with theirs.

### 4.4. The Accretion Rates

The distributions of the accretion rates of radio galaxies are shown in Figure 4. The distribution range of the accretion rates of radio galaxies is mainly from $10^{-7.5}$ to $10^{-3.5}$. Radio galaxies have low accretion rates. Sbarrato et al. (2014) also found that radio galaxies have $L_{BLR}/L_{Edd} < 5 \times 10^{-4}$. Wu & Cao (2008) have suggested that radio galaxies with low accretion rates ($\dot{m} < 0.01$) may be in the ADAF accretion mode. These results may imply that the radio galaxies with gamma-ray emission have radiatively inefficient accretion disks (i.e., ADAFs). The average values of the accretion rates of FR II radio galaxies and FR I radio galaxies are $\langle \log L_{\rm disk}/L_{\rm Edd} \rangle|_{\rm FRII} = -4.60 \pm 0.64$ and $\langle \log L_{\rm disk}/L_{\rm Edd} \rangle|_{\rm FRI} = -5.59 \pm 0.68$, respectively. Through a parameter T-test ($P = 0.0006$), nonparametric K-S test ($P = 0.003$), and Kruskal–Wallis H-test ($P = 0.002$), we find that the distributions of the accretion rates between FR II radio galaxies and FR I radio galaxies are significantly different.

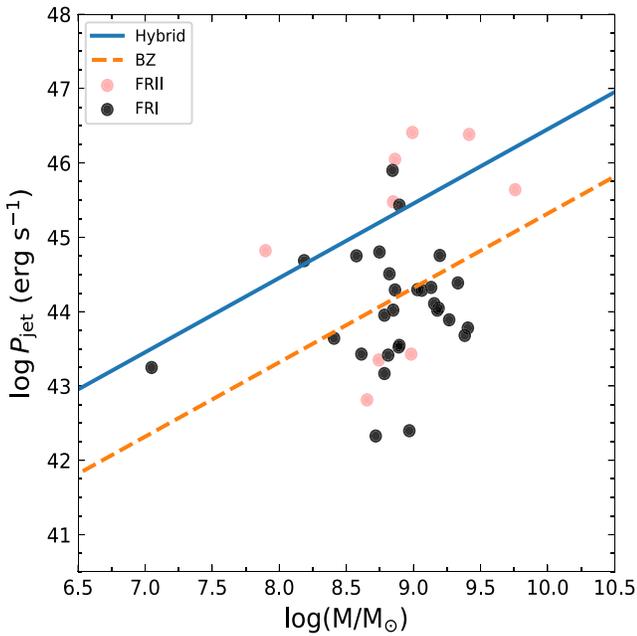

**Figure 5.** Relation between the jet kinetic power extracted from the underlying ADAFs for the BZ and hybrid models and the black hole mass, respectively. The $\dot{m} = 0.01$ and $\alpha = 0.3$ are adopted (Narayan & Yi 1995). The spin of the black hole $j = 0.98$ is adopted. The red dots show FR II radio galaxies and the black dots show FR I radio galaxies. The orange line is the BZ jet model, and the blue line is the hybrid model.

### 4.5. The Jet Kinetic Power versus Black Hole Mass

Figure 5 shows the relation between the jet power extracted from the ADAFs for the BZ and hybrid models and the black hole mass, respectively. The orange dashed line is the BZ jet model, and the blue line is the hybrid model. We find that the jet kinetic power of about 82% of radio galaxies can be explained by the hybrid model. Tombesi et al. (2010) have suggested that some radio galaxies have ultrafast outflows with $v \sim (0.04–0.15)c$. Although these winds move at relativistic speeds, their findings indicate that there are also some magnetohydrodynamic wind activities in AGNs with relativistic jets, thus strengthening the idea of a hybrid mechanism (Foschini 2011). Some authors have suggested that the jets of radio galaxies may be explained by the hybrid model with ADAFs surrounding spinning black holes (e.g., Wu et al. 2013; Feng & Wu 2017). Our results further confirm their conclusions. At the same time, we also find that the jet kinetic power of about 69% of FR I radio galaxies and 33% of FR II radio galaxies can be explained by the BZ jet model. Wu et al. (2011) have also proposed that the FR I radio galaxies are powered by fast-rotating BHs based on the BZ process. We also find that some high jet power sources cannot be explained by ADAFs. There are two possible explanations. One is that the accretion disks of these high jet power sources are not in an ADAF state. Yuan (2001) has suggested that the maximal jet power extracted from a spinning black hole surrounded by such a hot luminous disk may be higher than that for pure ADAFs. This may help to explain the high jet power in some radio galaxies. The other is that it needs to be explained by other jet models. Cao (2018) has suggested that the magnetic field being dragged inward by an accretion disk with magnetization-driven outflows may accelerate the jets (Cao & Spruit 2013). These high jet power sources may be explained by the accretion disk with magnetization-driven outflows model.

### 4.6. The Jet Power versus Accretion Disk Luminosity

It is generally believed that there is a close connection between jets and accretion. Many authors have confirmed this relationship (e.g., Rawlings & Saunders 1991; Cao & Jiang 1999; Ghisellini et al. 2010, 2014; Sbarrato et al. 2014; Chen et al. 2015, 2023). We also study this relationship for radio galaxies. The relations between jet radiation power (top panel) and jet kinetic power (bottom panel) and accretion disk luminosity for radio galaxies are shown in Figure 6. We find a weak correlation between jet radiation power and accretion disk luminosity for the whole sample ($r = 0.30$, $P = 0.06$). The tests of Spearman ($r = 0.41$, $P = 0.01$) and Kendall tau ($r = 0.29$, $P = 0.009$) show a significant correlation between jet radiation power and accretion disk luminosity for the whole sample. There is also a significant correlation between jet kinetic power and accretion disk luminosity for the whole sample ($r = 0.43$, $P = 0.006$). The tests of Spearman ($r = 0.41$, $P = 0.01$) and Kendall tau ($r = 0.29$, $P = 0.009$) also show a significant correlation between jet kinetic power and accretion disk luminosity for the whole sample. Rajguru & Chatterjee (2022) studied the relation between jet power and accretion disk luminosity using a large sample of Fermi blazars. They found a weak correlation after excluding the common redshift dependence. We also use partial correlation analysis to study this relationship and find that there is a weak correlation between jet kinetic power and accretion disk luminosity ($r = 0.32$, $P = 0.052$). However, there is always a strong correlation between the radiation jet power and the luminosity of the accretion disk, even if the dependence on redshift is excluded ($r = 0.43$, $P = 0.007$).

### 4.7. IC Luminosity versus Synchrotron Luminosity

The relation between IC luminosity and synchrotron luminosity for radio galaxies is shown in Figure 7. We find a significant Pearson correlation between them for the whole sample ($r = 0.74$, $P = 1.04 \times 10^{-7}$). The Spearman correlation coefficient and significance level are $r = 0.71$ and $P = 5.19 \times 10^{-7}$. The Kendall tau correlation coefficient and significance level are $r = 0.53$ and $P = 3.08 \times 10^{-6}$. These two tests also show a significant correlation. The symmetric least-squares linear regression gives the best linear fitting equation as $\log L_{\rm IC} = (1.07 \pm 0.16)\log L_{\rm syn} + (-3.92 \pm 7.11)$. According to Ghisellini (1996; $L_{\rm EC} \sim L_{\rm syn}^{1.5}$, $L_{\rm SSC} \sim L_{\rm syn}^{1.0}$), our results suggest that the IC components of radio galaxies are dominated by the SSC process. We find that the galaxies in our sample have low accretion rates ($L_{\rm disk}/L_{\rm Edd} < 10^{-2}$). The BL Lacs also have low accretion rates ($L_{\rm disk}/L_{\rm Edd} < 10^{-2}$; Sbarrato et al. 2014). Some authors have suggested that the high-energy components of BL Lacs are dominated by the SSC process (e.g., Celotti & Ghisellini 2008; Ghisellini et al. 2010; Lister et al. 2011; Ackermann et al. 2012; Zhang et al. 2013). These results may imply that the high-energy component of a source with a low accretion rate is dominated by the SSC process.

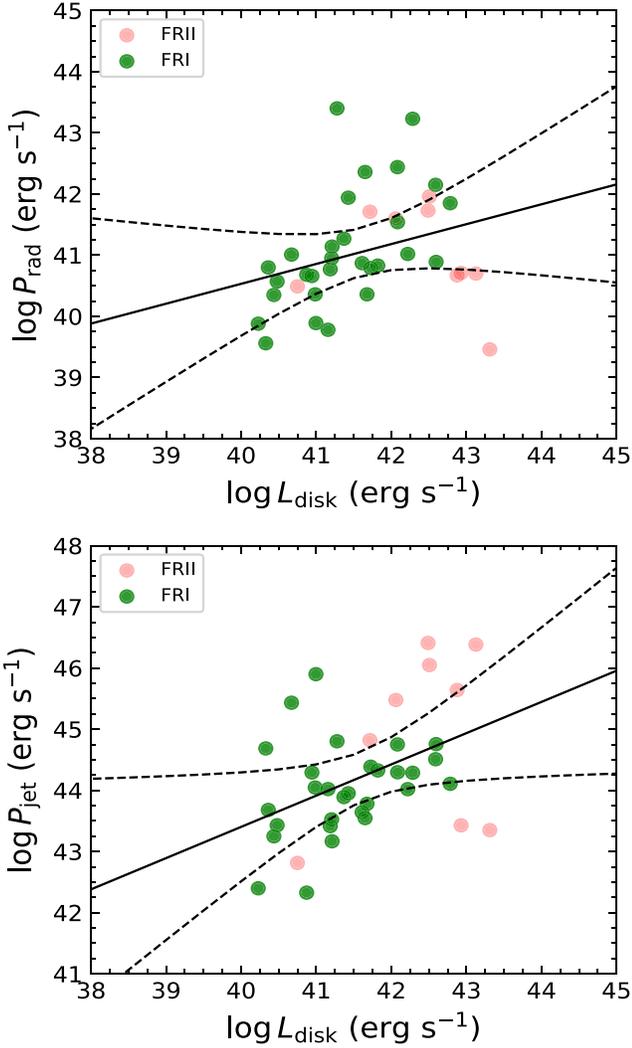

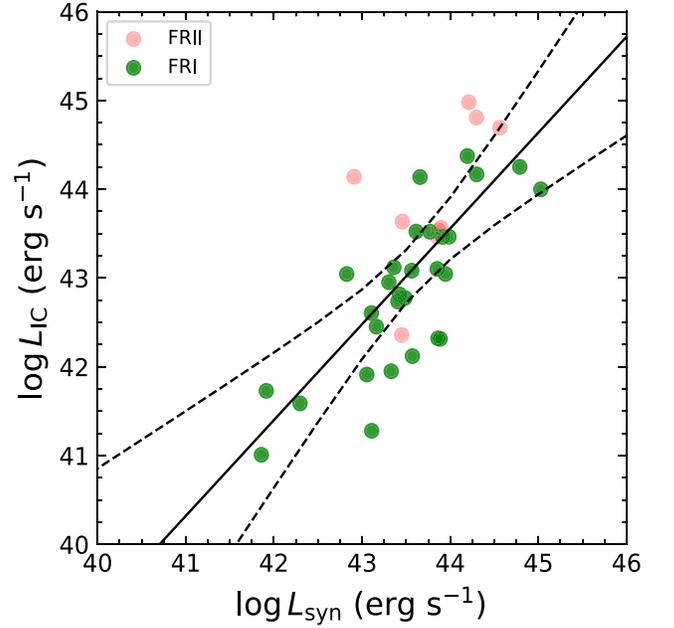

Figure 7. Relation between IC luminosity and synchrotron luminosity for radio galaxies. The red dots show FR II radio galaxies and the green dots show FR I radio galaxies. The solid line corresponds to the best-fitting linear model obtained with the symmetric least-squares fit. The dashed lines indicate $3\sigma$ confidence bands.

Figure 6. Relations between jet radiation power (top) and jet kinetic power (bottom), respectively, and accretion disk luminosity for radio galaxies. The red dots show FR II radio galaxies and the green dots show FR I radio galaxies. The solid lines correspond to the best-fitting linear models obtained with a symmetric least-squares fit. The dashed lines indicate $3\sigma$ confidence bands.
$\log P_{\rm rad} = (0.33 \pm 0.17)\log L_{\rm disk} + (27.53 \pm 7.07)$;
$\log P_{\rm jet} = (0.51 \pm 0.18)\log L_{\rm disk} + (22.97 \pm 7.42)$.

## 5. Conclusions

In this work, we have performed a broadband study of a large sample of radio galaxies with gamma-ray emission. Our main findings are as follows:

1. The average Doppler factors and magnetic fields of FR II radio galaxies and FR I radio galaxies are consistent with each other within the measurement errors. There are no significant differences between FR I radio galaxies and FR II radio galaxies in the distributions of the Doppler factors and magnetic fields.

2. Compared with FR I radio galaxies, FR II radio galaxies have similar average jet powers in the form of radiation ($P_{\rm rad}$), electrons ($P_{\rm e}$), cold protons ($P_{\rm p}$), magnetic fields ($P_{\rm mag}$), and kinetics ($P_{\rm jet}$). Through a parameter T-test, nonparametric K-S test, and Kruskal–Wallis H-test, we find that the distributions of these physical parameters between FR II radio galaxies and FR I radio galaxies are not significantly different.

3. The FR II radio galaxies have higher average accretion rates than the FR I radio galaxies. There is a significant difference between the FR II radio galaxies and FR I radio galaxies in the distribution of accretion rates.

4. We investigate the jet power of the BZ mechanism and hybrid mechanism based on self-similar solutions of ADFAs surrounding Kerr black holes. According to the relation between jet kinetic power and black hole mass, we find that the jet kinetic power of most radio galaxies can be explained by a hybrid model. The jet kinetic power of about 69% of FR I radio galaxies and 33% of FR II radio galaxies can be explained by the BZ mechanism.

5. There is a significant correlation between jet kinetic power and accretion disk luminosity for radio galaxies. However, the correlation between jet kinetic power and accretion disk luminosity shows a weak correlation when redshift is excluded. There is a strong correlation between the jet power of radiation and accretion disk luminosity, even after redshift has been excluded.

6. We find a significant correlation between IC luminosity and synchrotron luminosity for radio galaxies. The slope of the best linear regression between IC luminosity and synchrotron luminosity in radio galaxies is $1.07 \pm 0.16$, which suggests that the high-energy components of radio galaxies are dominated by the SSC process.

Y.C. is grateful for financial support from the National Natural Science Foundation of China (No. 12203028). This work was supported from the research project of Qujing Normal University (2105098001/094). This work is supported by the youth project of Yunnan Provincial Science and Technology Department (202101AU070146, 2103010006). Y.C. is grateful for funding for the training program for talents in Xingdian, Yunnan Province. Q.G. is supported by the National Natural Science Foundation of China (Nos. 11733002, 12121003, 12192220, and 12192222). We also acknowledge the science research grants from the China Manned Space Project with No. CMS-CSST-2021-A05. This work is supported by the National Natural Science Foundation of China (11733001 and U2031201).

### ORCID iDs

Yongyun Chen (陈永云)  https://orcid.org/0000-0001-5895-0189